\documentstyle[12pt,aasms4]{article}
\def \s{~\rm{s}}
\def \km{~\rm{km}}
\def \AU{~\rm{AU}}
\def \K{~\rm{K}}

\begin{document}


\title{
CAN PLANETS INFLUENCE THE  \\
HORIZONTAL BRANCH MORPHOLOGY?   }

\author{ 
Noam Soker }
\affil{
Department of Physics, University of Haifa at Oranim\\
Oranim, Tivon 36006, ISRAEL \\
soker@physics.technion.ac.il       }

\begin{abstract}
   
 I examine the influence of planets on the location of stars on the 
Hertzsprung-Russel diagram as the stars turn to the horizontal 
branch.
 As stars which have planetary systems evolve along the red giant branch 
and expand, they interact with the close planets, orbital separation 
of $\lesssim 5 \AU$. 
 The planets deposit angular momentum and energy
into the red giant stars' envelopes, both of which are likely to enhance
mass loss on the red giant branch. 
 The enhanced mass loss causes the star to become bluer as it turns to 
the horizontal branch.
 I propose that the presence of planetary systems, through this mechanism,
can explain some anomalies in horizontal branch morphologies. 
 In particular, planetary systems may be related to the 
``second parameter'', which determines the distribution of horizontal 
branch stars on the Hertzsprung-Russel diagram. 
 The distribution of planets' properties (e.g., mass, orbital separation, 
prevalence) in a specific globular cluster depends on several properties 
of the globular cluster itself (e.g, shape, density). 
 This dependence may explain some anomalies and variations in the 
horizontal branch morphologies between different globular clusters. 
 I estimate that in $\sim 40 \%$ of the cases the extreme horizontal 
branch stars may be formed from the influence of low mass stellar 
companions, making any prediction of the exact horizontal branch
morphologies very difficult. 
 The proposed scenario predicts that surviving massive planets or brown 
dwarfs orbit many of the extreme blue horizontal branch stars, 
at orbital periods of $\sim 10$ days. 

\end{abstract}



\section{INTRODUCTION}
 
 Sun-like stars which burn helium in their cores occupy the horizontal 
branch (HB) in the Hertzsprung-Russel (HR) diagram. 
 The distribution of stars on the HB of a stellar system is termed the 
HB morphology.  
 The HB morphologies differ substantially from one globular cluster to 
another. 
 It has long been known 
that metallicity is the main factor which determines the location
of HB stars on the HR diagram. 
 For more than 30 years, though, it has been clear that another factor is
required to explain the HB morphologies in different globular clusters
(Sandage \& Wildey 1967; van den Bergh 1967; see reviews by 
Fusi Pecci \& Bellazzini 1997; Rood  1997; Rood, Whitney, \& D'Cruz 1997). 
 This factor is termed the {\it second parameter} of the HB.  
 In recent years it has become clear that mass loss on the red giant branch
(RGB) is closely connected with the second parameter, in the sense that the
second parameter should determine the HB morphology by regulating
the mass loss on the RGB  (Dorman, Rood, \& O'Connell 1993; 
D'Cruz {\it et al.} 1996 and references therein). 
 On the RGB, which is the stage prior to the HB and before core helium 
ignition, the star is large and luminous and has a high mass loss rate.
 In the present study I suggest that in many cases interaction of RGB 
stars with brown dwarfs and gas giant planets regulates the stellar 
mass loss on the RGB. 
 In $\sim 40 \%$ of the cases the interaction, i.e., deposition of angular
momentum {{{  and energy}}}, is with a wide stellar companion.  

 That planets may play a significant role in the evolution of evolved
stars has been suggested before. 
Peterson, Tarbell \& Carney (1983) mentioned the possibility that planets 
can spin-up RGB stars, though later they abandoned this idea.
 In earlier studies (Soker 1996, 1997) 
I postulated that most elliptical planetary nebulae result from the 
influence of substellar objects, mainly gas giant planets, on the mass
loss geometry from asymptotic giant branch (AGB) stars. 
 I derived (Soker 1996) the maximal orbital separations 
allowed for brown dwarfs and massive planets in order to tidally spin-up 
progenitors of planetary nebulae, and found it to be $\sim 5~$AU.  
For a substellar object to have a high probability of being present 
within this orbital radius, on average several substellar objects must 
be present around most main sequence stars of masses $\lesssim 5 M_\odot$. 
 According to the star-planet interaction scenario, $\sim 50 \%$ of 
all main sequence stars which are progenitors of planetary nebulae
should have such planetary systems (Soker 1997). 
 It is important to note that the statistical analysis of Soker (1997) 
refers to 458 planetary nebulae, all in the field, and therefore the
conclusions may be different for globular clusters.  

  Soker (1998) shows that in most cases low mass stars interact with binary 
companions, stellar or substellar, already on the RGB, as opposed to
stars with masses of $M \gtrsim 2 M_\odot$, which interact with their 
companions mainly on the AGB. 
 I argue below that this interaction results in an enhanced mass loss 
which depends on the properties of the interacting planets, and hence 
can explain some anomalies and differences in HB morphologies if 
planetary systems' properties vary from one globular cluster to another.  

\section{THE SECOND PARAMETER PROBLEM}
 
 In several globular clusters there are bimodal distributions of red 
and blue HB stars (e.g., NGC 2808, Ferraro {\it et al.} 1990; NGC 1851,
Walker 1992; more examples in Catelan {\it et al.} 1997). 
 There are many differences between the HB morphologies in the different 
globular clusters, (e.g., Catelan {\it et al.} 1997), and these two 
clusters, for example, have different types of bimodal distributions.  
 In the present paper I do not attempt to explain all these fine details
of the HB morphologies, but I limit myself to point out the 
significant role that planets can play in determining the HB morphologies.  
 It is clear that not only planets play a role in the proposed second 
parameter mechanism due to companions, since, for example, stellar binary 
companions and occasional collisions with passing stars can also influence 
the mass loss on the RGB. 
 The distribution in the HR diagram requires that the extreme HB 
stars lose up to almost all their envelope while on the RGB 
(Dorman {\it et al.} 1993; D'Cruz {\it et al.} 1996). 
D'Cruz {\it et al.} (1996) show that they can reproduce the basic 
morphology of the HB in different globular clusters by assuming simple 
mass loss  behavior on the RGB.
 They could even produce a bimodal distribution in solar metallicity
stellar groups (e.g., open clusters).     
 However, there are still open questions. 
\newline
{\bf (1)}  The bimodal distribution is also found 
in globular clusters having metallicity much below solar. 
 It is not found in metal-poor globular clusters, however, 
in contrast to gaps, which are found at all metallicities 
(Catelan {\it et al.} 1997). 
\newline 
{\bf (2)} What determines the distribution of the mass 
loss rates on the RGB?
 The factor which determines this mass loss is probably the second parameter
(D'Cruz {\it et al.} 1996).
 Comparison of globular clusters which have similar properties 
but different HB morphologies with globular clusters which have similar 
HB morphologies but are different in specific properties (e.g., metallicity) 
led to no single factor (e.g., Ferraro {\it et al.} 1997 and references
therein). 
 The popular factors that were examined and found short of producing
the effect of the second parameter are age 
(Stetson, van den Bergh, \& Bolte 1996), globular cluster density
(Ferraro {\it et al.} 1997), metallicity, rotation  
(Peterson, Rood, \& Crocker 1995; but see Sweigart 1997a,b), 
and stellar collision and merger (Rich {\it et al.} 1997; Rood 1997).  
 Several authors (e.g., Peterson {\it et al.} 1995) suggest that two or
more factors acting together produce the second parameter. 
 However, they could not point either to the factors or to the process
by which they determine the HB morphology. 
In $\S 4$ I return to this question in the frame of 
the star-planet interaction model.
\newline 
{\bf (3)} Why are some HB stars observed to have projected rotation 
velocities as high as $\sim 40 \km \s^{-1}$ (Peterson {\it et al.} 1995; Cohen \& McCarthy 1997)?
Harpaz \& Soker (1994) show that the envelope's angular
momentum of evolved stars decreases with mass loss as 
$L_{\rm env} \propto M_{\rm env}^{3}$, where the envelope density 
distribution is taken as $\rho \propto r^{-2}$ and a solid body 
rotation is assumed to persist in the entire envelope. 
 Therefore, I do not expect HB stars, after losing more than 
$1/3$ of their envelope on the RGB, to rotate at such high velocities. 
 Indeed, to account for the fast rotating HB stars
Peterson {\it et al.} (1983) already mentioned the 
possibility that planets can spin-up RGB stars,  
although they abandoned the planet idea later. 
 They cite the amount of the required angular momentum to be
about equal to the orbital angular momentum of Jupiter, which is $\sim 100$
times larger than that of the sun today. 
  The reason to assume spin-up by planets is that single sun-like stars 
are likely to rotate very slowly when reaching the HB. 
 In addition to angular momentum loss on the RGB (Harpaz \& Soker 1994),
the star will not preserve much angular momentum from the main sequence.
On the observational side, Tomczyk, Schou, \& Thompson (1995) 
find that the sun rotates as a solid body down to $r \simeq 0.2 R_\odot$, 
so there is no storage of angular momentum in the solar interior.
 On the theoretical side, Balbus \& Hawley (1994) argue that the powerful 
weak-field MHD instability is likely to force a solid body rotation
in the radiative zone of stars. 
 Therefore, I think, and this is in dispute with some other studies 
(e.g., Pinsonneault, Deliyannis, \& Demarque 1991), 
that single low mass stars cannot store a substantial amount of angular 
momentum.
\newline
{\bf (4)} In a recent paper Sosin {\it et al.} (1997) show that 
in the globular cluster NGC 2808 there are three subgroups in the blue HB.  
 Similar subgroups were found in the globular cluster M13 
(Ferraro {\it et al.} 1997). 
 Based on the stellar evolutionary calculations of Dorman {\it et al.} (1993),
Sosin {\it et al.} (1997) claim that the subgroups on the blue HB of NGC 2808
correspond to stars having envelope masses of 
$M_{\rm {env}} \lesssim 0.01 M_\odot$, 
$0.02 \lesssim M_{\rm {env}} \lesssim 0.055 M_\odot$, 
and $0.065 \lesssim M_{\rm {env}} \lesssim 0.13 M_\odot$ from blue to red.
 The red HB stars are concentrated in the range  
$0.16 \lesssim M_{\rm {env}} \lesssim 0.22 M_\odot$.  
 The number of stars in each group, from red to blue, is 
$\sim 350$, 275, 70 and 60. 
What is the cause of these subgroups in NGC 2808? 
 
\section{PLANETS AND THE SECOND PARAMETER}
\subsection{Planets Versus Stellar Companions}
 
 Let us examine how interaction of RGB stars with planets and brown dwarfs 
can influence the HB morphology.  
 First, there are stars that will not interact on the RGB with any 
gas giant planet or other companions. 
 These stars will lose little mass, and they will form the red HB. 
 The blue HB stars, I suggest, result from RGB stars that interact
with gas giant planets, brown dwarfs, or stars.  
 {{{  The interaction with stellar companions is left to a future study.
 Among all interaction with companions, I estimate that in
$\sim 40 \%$ of the cases the interactions are with stellar companions, 
i.e.,  $\sim 2/3$ of the number of planetary systems and
brown dwarfs. 
 This estimate is based on the following considerations. 
The total fraction of PNe expected to contain close stellar binary 
systems at their centers is $\sim 22 \%$ of all PNe (Yungelson, 
Tutukov, \& Livio 1993), while Schwarz \& Corradi (1995) 
estimate that bipolar PNe constitute $\sim 11 \%$ of all PNe. 
 Similar numbers were obtained by Han, Podsiadlowski, \& Eggleton (1995)
in their Monte Carlo simulations. 
 Eggleton (1993) estimate that $\sim 20 \%$ of all systems are stellar 
binaries with periods of $\lesssim 10^4$ days. 
 In a previous paper (Soker 1997)  I assumed that most bipolar planetary 
nebulae result from companions which avoid the common envelope phase, 
and concluded, based on the numbers given above and the analysis in that
paper, that $\sim 35 \%$ of all planetary nebula progenitors interacted with 
stellar companions, as opposed to $\sim 55 \%$ that interacted with 
planetary systems.  
}}}

 A planet entering the envelope of a RGB star releases energy and
angular momentum, both of which are expected to increase the mass loss 
rate. 
 If the planet is evaporated near the core, the mass loss rate may decrease
for a very short time (Harpaz \& Soker 1994), but overall the 
planet increases the total mass that is lost on the RGB. 
Hence, the star reaches the HB with less mass in its envelope. 
  There are three evolutionary routes for star-planet systems 
(Livio \& Soker 1984): ($i$) evaporation of the planet in the envelope;
($ii$) collision of the planet with the core (i.e., the planet overflows 
its Roche lobe when at $\sim 1 R_\odot$ from the core); 
and ($iii$) expelling the envelope while the planet survives the 
common envelope evolution. 
  To show that these three routes may explain the three subgroups found by 
Sosin {\it et al.} (1997) in the blue HB of the globular cluster 2808,
I study the fate of the envelope and the planet after the common envelope
phase. 

\subsection{Accretion and Evaporation}
 It is not clear whether a planet inside an extended envelope accretes 
at the Bondi \& Hoyle (1994) accretion rate, as assumed 
by Livio \& Soker (1984), 
or whether it will expand, form a ``blanket'' of $<0.05 M_\odot$, and 
stop accreting, as was found by Hjellming \& Taam (1991) for a stellar 
secondary. 
{{{  Here I will consider the case without accretion.
Taking accretion into account, as in Livio \& Soker (1984), results
in evaporation as well (see figs. 3 and 5 by Livio \& Soker 1984). }}}
 Hjellming \& Taam (1991) find that 
at the end of the common envelope phase the secondary loses back 
to the common envelope most of the accreted mass. 
 Based on the calculations of Hjellming \& Taam (1991) 
I assume that only a small amount of mass ($< 0.1 M_p$) is accreted
by the planet, and that it forms a ``blanket'' extended to a radius of 
$\eta R_p$, where $M_p$ and $R_p$ are the mass and radius of the planet, 
respectively, and $\eta \lesssim 5$. 
 To find the approximate location of evaporation, I equate the
local sound speed in the RGB primary star's envelope 
to the escape velocity from the planet surface $v_e = (2 G M_p /R_p)^{1/2}$.
 {{{  This is justified by the results of Livio \& Soker (1984), who 
find evaporation of low mass planets at small radii.  
Using their expressions we can understand this as follows. 
 The evaporation rate is taken from Spitzer (1947), but the virial 
temperature is expressed in terms of the escape velocity from the planet and
the sound speed, 
\begin{eqnarray}
(\dot M_p)_{\rm {evap}} \simeq 
4 (\pi/6)^{1/2} C_s \rho_p R_p^2 
\left(1+{{\gamma V_e^2}\over{2 C_s^2}} \right) 
\exp \left(-{{\gamma V_e^2}\over{2 C_s^2}} \right) ,
\end{eqnarray}
where $C_s$ is the sound speed in the planet's atmosphere, $\rho_p$ is
the effective density in the escaping region, and $\gamma$ is the
adiabatic index. 
 Significant evaporation occurs when the exponent argument becomes
unity, i.e., $\gamma V_e^2 \simeq 2 C_s^2$. 
 To show that the planet's temperature $T_p$ cannot be too different from
the stellar temperature at the planet's location we write the
expression for the planet's luminosity (Livio \& Soker eq. 4)
\begin{eqnarray}
L_p \simeq 16 \pi \sigma R_p^2 (T_p^4-T^4)(4+3 \beta \kappa \rho_p R_p)^{-1},
\end{eqnarray}
where $\beta$ is a parameter in the range $0.1-1$, and $\kappa$ is
the opacity.
 Due to the strong dependence on the temperature, if $T \gg T_p$ the
radiation flux from the surroundings of the planet will rapidly
heat it.  
 Even when the planet is allowed to accrete its temperature is close
to the surrounding temperature, and eventually low mass planets are
evaporated (Livio \& Soker 1984). 
 If the planet survives the evolution until the star turns to the HB,
the radiation from the star is unable to evaporate the planet. 
}}}

 For the radii of brown dwarfs and gas giant planets I take 
$R_p \simeq 0.1 R_\odot$. 
%
 The temperature inside the convective envelope of a late RGB 
star can be approximated by $T \simeq 2 \times 10^6 (r/R_\odot)^{-1} \K$ 
(Harpaz 1998), similar to that of AGB stars (Soker 1992),
where $r$ is the distance from the center of the star. 
 Equating the envelope's sound speed to the planet's escape velocity 
gives the approximate location of evaporation
\begin{eqnarray}
a_{\rm {EVA}} \simeq 10 \left( \frac {M_p}{M_J} \right)^{-1} R_\odot,
\end{eqnarray}
where $M_J=0.001 M_\odot$ is Jupiter's mass. 
 The planet's destruction will occur when its radius exceeds
the radius of the Roche potential. 
 Mass transfer from the planet to the primary's core will start when 
the blanket radius exceeds the Roche radius.  
 The radius of the Roche lobe of a low mass secondary is
$R_{\rm {RL}} \simeq 0.46 a (M_p/M_c)^{1/3}$ (Paczynski 1967),   
where $a$ is the orbital separation and $M_c$ is the primary's core mass. 
 For stars on the tip of the RGB we can take 
$M_c \simeq 0.5 M_\odot \simeq 500 M_J$. 
 For the planet's radius we take $0.1 \eta R_\odot$. 
 The orbital separation at which Roche lobe overflow starts is 
\begin{eqnarray}
a_{\rm {RLO}} \simeq 1.7 \eta 
\left( \frac {M_p}{M_J} \right)^{-1/3} R_\odot.
\end{eqnarray}

\subsection{Spinning-up the Stellar Envelope}

 When the star evolves along the RGB it expands slowly.
 When its radius $R$ becomes $\sim 20 \%$ of the orbital separation $a_0$, 
tidal forces will cause the substellar companion orbit to decay in a time 
shorter than the evolutionary time (Soker 1996), thus forming a common 
envelope phase. 
  As it spirals inside the envelope, the planet (or any other companion)
deposits energy and angular momentum. 
 The angular velocity of the envelope $\omega$ can be estimated as follows. 
Approximating the envelope's density profile as $\rho \propto r^{-2}$
(Soker 1992; Harpaz 1998), we find the envelope's moment of inertia to be 
$I_{\rm {env}}= 2 M_{\rm {env}} R^2/9$. 
 The final envelope angular momentum $I \omega$ is equal to the planet's  
initial orbital angular momentum 
$M_p (G M_1 a_0)^{1/2} = M_p \omega_{\rm {Kep}} (a_0R^3)^{1/2}$,
where $\omega_{\rm {Kep}}$ is the Keplerian angular velocity on the 
RGB star's surface, and $M_1$ is the primary's total mass. 
 Substituting $a_0 = 5R$, as discussed above, we find for the envelope 
angular velocity
\begin{eqnarray}
\frac{\omega}{\omega_{\rm {Kep}}}  
= 0.03 
\left( \frac{M_p}{M_J} \right)  
\left( \frac{M_{\rm {env}}}{0.3 M_\odot} \right) ^{-1} .
\end{eqnarray}
 Wide stellar companions  ($2 \AU \lesssim$$a_0 \lesssim 20 \AU$) can deposit angular 
momentum via tidal interaction, leading to similar effects as those of 
planets.  
 This idea is supported by the recent finding that some blue HB stars 
have wide stellar companions (Liebert 1997).   
 Sweigart (1997a, b) suggests that rotation can lead to the mixing of helium 
from the core to the envelope on the RGB.   
 This increases the RGB tip luminosity, and hence total mass loss on
the RGB, leading to the formation of blue HB stars. 
 Sweigart (1997a, b) suggests that this can explain the second parameter, 
though he does not mention the required angular velocity and how his model
accounts for the different groups on the HB.   
 The deposition of angular momentum by planets may put the scenario
of helium mixing on more solid ground.

\subsection{The Fate of the Planets}
 
 For a final orbital separation $a_f \ll a_0$, and a substellar companion, 
the envelope mass removed by the gravitational energy of the star-planet 
system $\Delta M_{\rm {env}}$
is given by (e.g., Iben \& Livio 1993) 
\begin{eqnarray}
\alpha \frac{G M_p M_c}{a_f} \simeq \frac{G M_c \Delta M_{\rm {env}}}{R}.
\end{eqnarray}
The parameter $\alpha$ is the efficiency of envelope removal:
part of the gravitational energy released will be channeled into other forms
rather than envelope removal; this will reduce $\alpha$.
By changing the properties of the envelope (e.g., excitation of non-radial 
pressure modes, Soker 1997; spinning-up the envelope) the companion
can further increase mass loss; these effects increase $\alpha$
{{{  (see discussion by Livio \& Soker 1988). }}}

 First consider planets that are evaporated before the envelope is lost.
 Taking for the final radius in equation (6) the evaporation radius 
from equation (3), we find for the envelope mass 
removed by the planet's energy deposition  
\begin{eqnarray}
\frac{\Delta M_{\rm {env}}}{0.1 M_\odot} \simeq
0.1 \alpha 
\left( \frac{M_p}{M_J} \right)^{2}  
\left( \frac{R}{100 R_\odot} \right). 
\end{eqnarray}
 We note that this equation refers only to the energy released while the
secondary spirals-in inside the envelope. 
 However, even a Jupiter-like planet spins the envelope to $\sim 1\%$
of the Keplerian velocity, and can cause the mass loss rate on the
RGB to increase above its value for non-rotating stars. 
 From equation (5) it seems that direct effects 
due to spinning-up (i.e., centrifugal force)
on the mass loss rate are significant for $M_p \gtrsim 3 M_J$. 
 I expect that non-direct effects of rotation (e.g., helium mixing, 
Sweigart 1997a; excitation of p-modes, Soker 1997) may increase the 
mass loss rate even for planets of only several$\times 0.1 M_J$.
 When the RGB star turns into an HB star, it has envelope mass lower than
that of non-interacting stars with the same initial properties, 
and it rotates somewhat faster. 
 
 Planets with small initial orbital separation have less angular momentum 
and enter the star's envelope at early RGB phases. 
The star will spend a longer time on the RGB after the planet's evaporation, 
and hence will lose more of its angular momentum. 
 These stars will, on average and depending on the planet mass, 
be slowly rotating blue HB stars.   
 Planets with larger initial orbital separation enter the envelope at
late RGB phases. 
 The star has already lost some mass before the interaction, and it will
lose less mass after the interaction than in the previous case
(for the same planet mass).  
 This star will reach the HB with more angular momentum than in the 
previous case (of interaction early on the RGB).
 This shows that a wide range of angular momentum can result from the
same initial planet masses.
 The exact location on the HB depends more on the planet's mass
than on its initial orbital separation, while the final angular 
momentum depends more on the initial orbital separation, 
for the reason discussed above.
 Hence, no clear correlation is expected between the location 
on the blue HB and the angular momentum on the HB, according to the
proposed planetary interaction. 
{{{  This discussion does not necessarily hold for interaction with
stellar companions, which is beyond the scope of this paper.}}} 
 Red HB stars, though, are expected to rotate slowly. 
 
 The condition for planets to survive is that the entire envelope be lost
before they are evaporated or overflow their Roche lobes. 
Taking $\Delta M_{\rm {env}}=M_{\rm {env}}$ in equations (6), we find 
for the final orbital separation of a surviving planet  
\begin{eqnarray}
{a_f} = 3 \alpha 
\left( \frac{M_p}{10M_J} \right)  
\left( \frac{R}{100 R_\odot} \right) 
\left( \frac{M_{\rm {env}}}{0.3 M_\odot} \right)^{-1} R_\odot. 
\end{eqnarray}
 Comparing this to equation (3) for the evaporation radius, we find
that in order not to be evaporated we require  $M_p \gtrsim 5 M_J$. 
 Considering the many uncertainties in the evaporation process and the 
values of the physical parameters (e.g., $\alpha$, $R$, $M_{\rm {env}}$),
the minimum planet mass required to survive evaporation can be in the range 
$\sim 1 M_J - 10 M_J$. 
 
 Surviving low mass planets may overflow their Roche lobes. 
 The condition for not overflowing the Roche lobe is $a_f > a_{\rm {RLO}}$,
where $a_{\rm {RLO}}$ is given by equation (4). 
 Using equations (4) and (8), and taking the initial orbital separation to 
be $a_0=5R$, as we did when deriving equation (5), 
we find the condition for planets not to overflow their Roche lobe to be
\begin{eqnarray}
\left( \frac{M_p}{10M_J} \right)^{4/3}  
\left( \frac{a_0}{500 R_\odot} \right) 
\gtrsim  0.3  \frac{\eta}{\alpha} 
\left( \frac{M_{\rm {env}}}{0.3 M_\odot} \right). 
\end{eqnarray}
 
 Although equations (8) and (9) contain several poorly determined
parameters (e.g., $\eta$, $\alpha$), physical variables (e.g., 
the initial orbital separation and the primary's radius at the time 
the planets enter the common envelope), and physical processes (e.g., tidal 
interaction), we can draw very interesting conclusions from these equations. 
  It turns out that there is a range in planet masses (which depends on the
quantities listed above) for which the planets overflow their Roche lobes 
before the entire envelope is expelled, and before the planets are evaporated.
 This range is $1 M_J \lesssim M_p \lesssim 10 M_J$, again, depending on the
initial orbital separation and other variables. 
 These planets will expel most of the envelope, but not all of it, when
overflowing their Roche lobe. 
 Planet matter leaving the Roche lobe will flow toward the core and will
further release gravitational energy. 
 The cool planet material can absorb heat and cause the star's radius 
and luminosity to decrease for a very short time 
(Harpaz \& Soker 1994). 
 The star recovers its initial structure on a dynamical time scale.
 The entire process of Roche lobe overflow and planet destruction can
take the star out of equilibrium for $\sim 10^2-10^3$ years.
This dynamical change can result in further mass being lost. 

\section{FROM PLANETARY SYSTEMS TO HB MORPHOLOGIES}

 Explaining the rich variety of HB morphologies requires an extensive
study, or rather several studies. 
 This is beyond the scope of this paper. 
 I therefore limit myself to applying the results of the previous section
to the HB morphology of NGC 2808 (Sosin {\it et al.} 1997).  
 Red HB stars, those that retain most of their initial envelope mass
(Dorman {\it et al.} 1993), are stars that did not interact with
gas giant planets, brown dwarfs, or stellar companions.  
 The blue HB stars, which lose more of their envelope mass on the RGB
(Dorman {\it et al.} 1993), result from stars that interact with
substellar companions while on the RGB. 
 Similar effects, of spin-up and enhanced mass loss rate, can result from 
tidal interaction with {\it stellar} companions which avoid a common 
envelope or enter a common envelope at very late stages of the RGB. 
{{{   In $\S 3.1$ I estimated the fraction of RGB stars interacting with 
stellar companions to be $\sim 40 \%$ of all interacting RGB stars. }}}

 The bluest subgroup, of very little mass, results from planets or brown 
dwarfs that expel most of the envelope and survive the common envelope.  
 They have mass of $M_p \gtrsim 10 M_J$. 
 They accrete a small amount of mass during the common envelope phase, 
which is lost back to the core of the star at the end of the common 
envelope phase. 
 These systems will form extreme blue HB stars with very low 
mass envelopes, $\lesssim 0.01 M_\odot$. 
 The envelope mass results from the mass lost by the surviving planet 
and a small amount that has been left in the inner regions of the 
original envelope. 

 The intermediate subgroup on the blue HB results from planets
that collide with the core. 
 Their masses are in the range $1 M_J \lesssim M_p \lesssim 10 M_J$.
 They expel much of the envelope, but not all of it. 
 The qualitative difference between the surviving and colliding planets
may explain the bluer gap near envelope mass of $\sim 0.01 M_\odot$. 
 The final Roche lobe overflow results in further release of gravitational
energy, and causes the envelope to shrink and expand again on a short
time scale. 
 Therefore, there is a ``gap'' between evaporated planets and planets 
that overflow their Roche lobes, in the sense that the latter group
release extra amounts of energy and cause envelope motion, both of which 
can lead to extra mass being lost from the envelope. 
 This, I suggest, may explain the HB gap near envelope mass of 
$0.06 M_\odot$ found by Sosin {\it et al.} (1997; see $\S 2$ above). 
 Lower mass planets, $M_p \lesssim 1 M_J$, are evaporated before expelling 
the envelope.
 Before being evaporated they release gravitational energy and spin-up the
star's envelope, both of which increase the mass loss rate on the RGB.
According to the proposed model, these planetary systems form the red 
side of the blue HB, 
{{{  or part of it, if stellar companions can also form these HB stars. 
This should be examined in a more extended study of interacting RGB stars.}}}

 Stellar binary mergers were suggested to account for the blue HB,
but Landsman {\it et al.} (1996) and Rood (1997) criticize this idea. 
 Rood's three comments against the {\it stellar}-binary scenario
do not hold against the star-planet scenario proposed here: 
(1) Planets do not change the general nature of the star (besides the mass
loss rate), contrary to stellar companions which collide with the star. 
(2) I do not expect strong variation with location in the cluster,  
unlike in scenarios with binary collisions (Rich {\it et al.} 1997).  
(3) I do not expect the star-planet interaction to strongly depend 
on the density of the cluster, unlike the case for stellar collisions. 
{{{   Some dependence on the location in the cluster and
cluster's density is expected, though. 
First, some fraction of the blue HB stars are expected to be formed from 
stellar companions or collisions ($\S 3.1$).
Second, the density of the cluster may affect the formation and
surviving of planetary systems. 
Third, stellar perturbations in the cluster center may 
cause the orbits of plants to shrink, increasing their changes to interact
with the RGB star, or the perturbations can destroy the planetary system
altogether. }}}

\section{FROM GLOBAL PROPERTIES TO PLANETARY SYSTEMS}

 How does the planetary system scenario account for the different HB 
morphologies of different globular clusters?
 The different morphologies result both from the efficiency 
of formation and the properties of planetary systems 
and from the evolution of stars on the RGB. 
 Since there is no basic theory to predict the properties of planetary
systems from initial conditions, I can only speculate on the global 
globular cluster properties which may determine the properties 
and formation of planetary systems. 
\newline
{\bf (1) Metallicity:} (a) Metallicity influences the efficiency of 
planet formation. 
There is no good theory to predict the efficiency, but low
metallicity results in lower efficiency. 
 On the other hand, in a globular cluster the HB stars result from 
main sequence stars fainter than the sun. 
 Fainter central stars evaporate the pre-planetary disk less, and
hence may allow Jovian planets to form more easily and closer to the star. 
(b) Metallicity determines the maximum radius which stars attain on the RGB,
being larger for metal-rich stars. 
 Larger radii increase the chances of interaction with planets. 
\newline
{\bf (2) Global cluster properties:} The global properties of the cluster
(e.g., shape, density of stars, initial mass function) may determine 
the efficiency of planet formation. 
The globular clusters M13 and M3, for example, have many similar 
properties, but  M13 is more elliptical than M3. 
 M3 has no blue HB, while M13 has an extended blue HB. 
 What is interesting for the star-planet interaction scenario is that
there are more blue stragglers in M3 than in M13 (Ferraro {\it et al.} 1997).
 This suggests that there are fewer stellar binary systems in M13. 
I would expect that if less stellar binary companions are formed,
then more planetary systems will form.  
 This might explain the anti-correlation of population on the
blue HB and the number of blue straggler stars observed in these
two globular clusters. 
\newline
{\bf (3) Age:} Age determines the initial mass (main sequence mass)
of the stars. 
 In addition, age influences both the envelope mass on the RGB, and 
the maximum radius on the RGB.
For that effect to be of any significance the age difference must be large. 
A large age difference by itself will have a strong effect on the basic 
HB morphology. 
 As mentioned above, the main sequence mass may determine the efficiency of 
planet formation as well. 
\newline
{\bf (4) Central cluster concentration:} 
This can influence the formation of planets (in an unknown way),
and lead to encounters with stars which may change the
orbits of the planets: destroying the planetary system or
shrinking the planets' orbits. 
 However, at the same time higher concentration will increase substantially
the cases of enhanced mass loss by stellar companions or passing stars. 
{{{  Therefore, it is possible that in cases where central density  
correlates with blue HB, stellar companions play the major role.}}}

\section{SUMMARY}
 The main points raised in the papers can be summarized as follows.
\newline 
(1) Close planets around low mass stars will interact with the
envelope as the stars evolve along the RGB. 
This will spin-up the star. 
 This may explain the fast rotation of blue HB stars.
 Because of angular momentum loss on the RGB there will be no strong 
correlation between the location on the blue HB and the rotational velocity.  
\newline
(2) The deposition of angular momentum and gravitational energy 
enhances the total mass lost on the RGB. 
As shown, e.g., by Dorman {\it et al.} (1993), as a result of the
mass loss the star will turn into a bluer HB star. 
\newline
(3) A planet orbiting inside an extended envelope can end in three ways:
evaporation, collision with the core, or survival while most of the
envelope is lost. 
 The three evolutionary routes may lead to concentration of stars on the
HB, i.e., gaps on the HB. 
As an example I proposed an explanation for the three subgroups on
the blue HB of the globular cluster NGC 2808. 
\newline
(4) Direct predictions are hard to make for a few reasons.  
First, there are several poorly known parameters (e.g., $\eta$, $\alpha$).
Second, in $\S 3.1$ I estimated that in $\sim 40 \%$ of the cases
stellar binary companions, rather than planets, will enhance the mass loss. 
Third, not much is know about the dependence of planets' formation
on global properties of clusters. 
Fourth, the influence of rotation on mass loss on the RGB
is poorly known.
\newline
(5) I argue that after the primary role of metallicity
(the first parameter of the HB morphology), 
planetary systems may play a significant role in determining the HB 
morphology.
 Therefore, the so called ``second parameter'' may be strongly connected 
to the presence of planetary systems in globular clusters. 
  However, I do not claim that the presence of planetary systems 
is the {\it only} agent affecting the second parameter. 
 Other stellar companions, collisions, rotation of single stars 
(Sweigart 1997 a, b) and as yet undetermined other factors may also
contribute to this as-yet unsolved problem.

\acknowledgments
This research has been supported in part by a grant from the University
of Haifa and a grant from the Israel Science Foundation.


\begin{references}

\reference{} Balbus, A. S., \& Hawley, J. F. 1994, MNRAS, 266, 769. 

\reference{} Bondi, H., \& Hoyle, F. 1944, MNRAS, 104, 273.

\reference{} Catelan, M., Borissova, J., Sweigart, A. V., \& Spassova N.
1997, ApJ, in press.  

\reference{} Cohen, J. G., \& McCarthy J. K. 1997, AJ, in press.

\reference{} D'Cruz, N. L., Dorman, B., Rood, R. T., \& O'Connell, R. W.
1996, ApJ, 466, 359. 

\reference{} Dorman, B., Rood, R. T., \& O'Connell, R. W. 1993, ApJ, 419, 596. 

\reference{} {{{  Eggleton, P. P. 1993,
{Astron. Soc. of the Pacific Conference Series}, 
Vol. 45, ed. D. D. Sasselov, p. 213. }}}


\reference{} Ferraro, F. R., Clementini, G., Fusi Pecci, F., 
Buonanno, R., \& Alcaino, G. 1990, A\&AS, 84, 59. 

\reference{} Ferraro, F. R., Paltrinieri, B., Fusi Pecci, F., Cacciari, C.,
Dorman, B. \& Rood,  R. T. 1997, ApJ, 484, L145.  

\reference{} Fusi Pecci, F.  \& Bellazzini, M. 1997,    
in The Third Conference on Faint Blue Stars, 
ed. A. G. D. Philip, J. Liebert, and R. A. Saffer 
in press. 

\reference{} {{{  Han, Z., Podsiadlowski, P., \& Eggleton P. P. 1995,
MNRAS, 272, 800.    }}}

\reference{} {{{  Harpaz, A. 1998, private communication. }}}

\reference{} Harpaz, A., \& Soker, N. 1994, MNRAS, 270, 734. 

\reference{} Hjellming, M. S., \& Taam, R. E. 1991, ApJ, 370, 709. 

\reference{} Iben, I. Jr., \& Livio, M. 1993, PASP, 105, 1373. 

\reference{} Landsman, W. B. {\it et al. }
1996, ApJ, 472, L93.  

\reference{} Liebert, J. 1997, private communication.  

\reference{} Livio, M., \& Soker, N. 1984, MNRAS, 208, 763.   

\reference{} {{{  Livio, M., \& Soker, N. 1988, ApJ, 329, 764.}}}

\reference{} Paczynski, B. 1967, Acta Astronomica, 17, 287.

\reference{} Peterson, R. C., Rood, R. T., \& Crocker, D. A. 1995, 
ApJ, 453, 214. 

\reference{} Peterson, R. C., Tarbell, T. D., \& Carney, B. W. 1983, 
ApJ, 265, 972.

\reference{} Pinsonneault, M. H., Deliyannis, C. P. \& Demarque, P. 
1991, ApJ, 367, 239.  

\reference{} Rich, R. M., {\it et al.} 1997, ApJ, 484, L25. 
 
\reference{} Rood, R. T. 1997,  in Fundamental Stellar Properties: The
Interaction Between Theory and Observation, eds. T. R. Bedding,
A. J. Booth, \& J. Davis (Dordrecht: Kluwer), in press. 

\reference{} Rood, R. T., Whitney, J., \& D'Cruz, N. 1997, 
in Advances in Stellar Evolution, eds. R. T. Rood \& A. Renzini
(Cambridge: Cambridge U. Press), in press. 

\reference{} Sandage, A. R. \& Wildey 1967, ApJ, 150, 469.  

\reference{} {{{  Schwarz, H. E., \& Corradi, R. L. M. 1995, 
Ann. of the Israel Physical Society, Vol. 11: 
{Asymmetrical Planetary Nebulae}, 
eds. A. Harpaz and N. Soker (Haifa, Israel), p. 113. }}}

\reference{} Soker, N. 1992, ApJ, 389, 628. 


\reference{} Soker, N. 1996, ApJL, 460, L53. 

\reference{} Soker, N. 1997,  ApJ Supp., 112, 487. 

\reference{} Soker, N. 1998,  ApJ, in press.   

\reference{} Sosin, C., {\it et al.} 1997, ApJ, 480, L35. 
 
\reference{} Spitzer, L. 1947, The Atmospheres of the Earth and Planets,
ed. Kuiper, G. P., University of Chicago.  

\reference{} Stetson, P. B., van den Bergh, S., \& Bolte, M. 1996, 
PASP, 108, 560. 

\reference{} Sweigart, A. V. 1997a, ApJ, 474, L23. 

\reference{} Sweigart, A. V. 1997b,  
in The Third Conference on Faint Blue Stars, ed. A. G. D. Philip, 
J. Liebert, and R. A. Saffer, 
in press 

\reference{} Tomczyk, S., Schou, J., \& Thompson, M. J. 1995, 
ApJL., 448, L57.

\reference{} van den Bergh, S. 1967, AJ, 72, 70.  

\reference{} Walker, A. R. 1992, PASP, 94, 1063. 

\reference{} {{{  Yungelson, L. R., Tutukov, A. V., \& Livio, M. 1993, 
ApJ, 418, 794.       }}}

\end{references}
\end{document}